\documentclass{webofc}
\usepackage[varg]{txfonts}   % Web of Conferences font
\usepackage{hyperref}
\usepackage{url}

\usepackage{lineno}
\usepackage{lipsum}
\hypersetup{colorlinks=true,citecolor=blue,urlcolor=blue,linkcolor=blue}
\newcommand{\pt}{$p_{\rm{T}}$}
\newcommand{\pte}{$p_{\rm{T,e}}$}

\newcommand{\mee}{$m_{\rm{ee}}$}

\newcommand{\gevc}{GeV/$c$}

\newcommand{\gevcc}{GeV/$c^{\rm 2}$}
\newcommand{\DCA}{DCA}

\newcommand{\DCAee}{DCA$_{\rm{ee}}$}

\newcommand{\DCAeez}{DCA$^{z}_{\rm{ee}}$}
\newcommand{\sqrtspp}{$\sqrt{s}$ = 13.6 TeV}
\newcommand{\sqrtsPbPb}{$\sqrt{s_{\rm{NN}}}$ = 5.36 TeV}

\begin{document}
\title{Dielectron production in pp and Pb--Pb collisions with \mbox{ALICE} in Run 3}

\author{\firstname{Florian} \lastname{Eisenhut for the ALICE Collaboration}\inst{1}\fnsep\thanks{\email{eisenhut@ikf.uni-frankfurt.de}}
}

\institute{Johann Wolfgang Goethe-University Frankfurt am Main, Germany}

\abstract{
The measurement of dielectron production is a fundamental piece of the puzzle in the understanding of the hot and dense matter produced in ultra-relativistic heavy-ion collisions. The dielectron spectrum provides information that penetrates the veil of final-state hadronic interactions and gives direct access to the early phases of the collision. However, the interpretation of the measured spectra relies on a precise understanding of all the contributing sources.

The measurement of dielectron production in proton--proton collisions, collected with the upgraded ALICE detector at \sqrtspp~is presented together with the status of the Pb--Pb analysis at \sqrtsPbPb. In particular, the extraction of prompt (and non-prompt) dielectron spectra over a wide mass range is explained. In addition it is discussed, in which way such analysis could help to understand the Drell--Yan process in a non-perturbative regime and to investigate the onset of thermal radiation.
}
\maketitle
\section{Introduction}
\label{intro}

Measuring dielectrons is an unique way of accessing the properties of the hot and dense medium (the quark-gluon plasma (QGP)) created in heavy-ion collisions. As electromagnetic probes, dielectrons are unaffected by the surrounding strongly interacting medium, and thus carry undistorted information about this medium at the time of their production. Unlike for photons, the invariant mass of dielectrons (\mee), which is correlated to the production time, can be used to study the evolution of the medium without being affected by its expansion. 
High-mass dielectrons (\mee~> 3.1 \gevcc) are produced in the early stages of the collisions, where hard scattering (e.g. the Drell--Yan process) and pre-equilibrium processes play a  significant role.
In the intermediate-mass region (IMR), i.e. 1.1 < \mee~ < 2.7 \gevcc, the production of thermal dielectrons form the QGP exceeds the one from hadronic matter.
In the late stages of the collisions primarily low-mass dielectrons (\mee~< 1.1 \gevcc) are emitted. This ordering allows to study different production mechanisms. 
However, such signals are very difficult to access due to the large combinatorial and physical background from hadronic decays. At LHC energies in particular, the physical background from correlated semileptonic decays of heavy-flavour (HF) hadrons dominates the IMR. Therefore, it is crucial to properly understand and handle the HF contribution.\linebreak
Minimum-bias (MB) proton--proton (pp) collisions are considered to provide a medium-free system and thus serve as an important baseline for studies of heavy-ion collisions. 
pp collisions provide an opportunity to study the contribution of the Drell--Yan process at low masses (\mee < 2.7 \gevcc), where pQCD calculations are unreliable due to large uncertainties.
%Finally, in high-multiplicity pp collisions the onset of possible thermal radiation can be investigated, as well as the search for new physics, such as dark photons.

\section{Detector upgrades and analysed data}
\label{Upgrades_Data}

In preparation for LHC Run 3, several of the ALICE apparatus's detectors were upgraded. 
The new GEM-based readout chambers of the time projection chamber (TPC) allow the readout rate to be increased by approximately a factor of 1000 for pp collisions and 100 for Pb--Pb collisions \cite{UpgradesLS2}. 
Furthermore, the new inner tracking system 2 (ITS2) based on CMOS MAPS technology, which has its first layer even closer to the beam axis than before (at 22 mm), provides an improved pointing resolution \cite{ITS2Performance}.
For the results presented, a sample of 277$\times 10^{9}$ (57.9$\times10^9$) MB pp events at \sqrtspp~from 2023 (2022) was analysed, which corresponds to an integrated luminosity of approximately \mbox{4.17 pb$^{-1}$} (0.97 pb$^{-1}$).
In total, this is a 600 times larger data sample than the one of the published dielectron result in Run 2 \cite{Run2pp13TeV}. 
For Pb--Pb collisions at \sqrtsPbPb~ an integrated luminosity of 3 nb$^{-1}$ has been recorded during the data taking period of 2023 and 2024.

\section{Results}
\label{results}

\begin{figure}[h]
\centering
\includegraphics[width=5cm,clip]{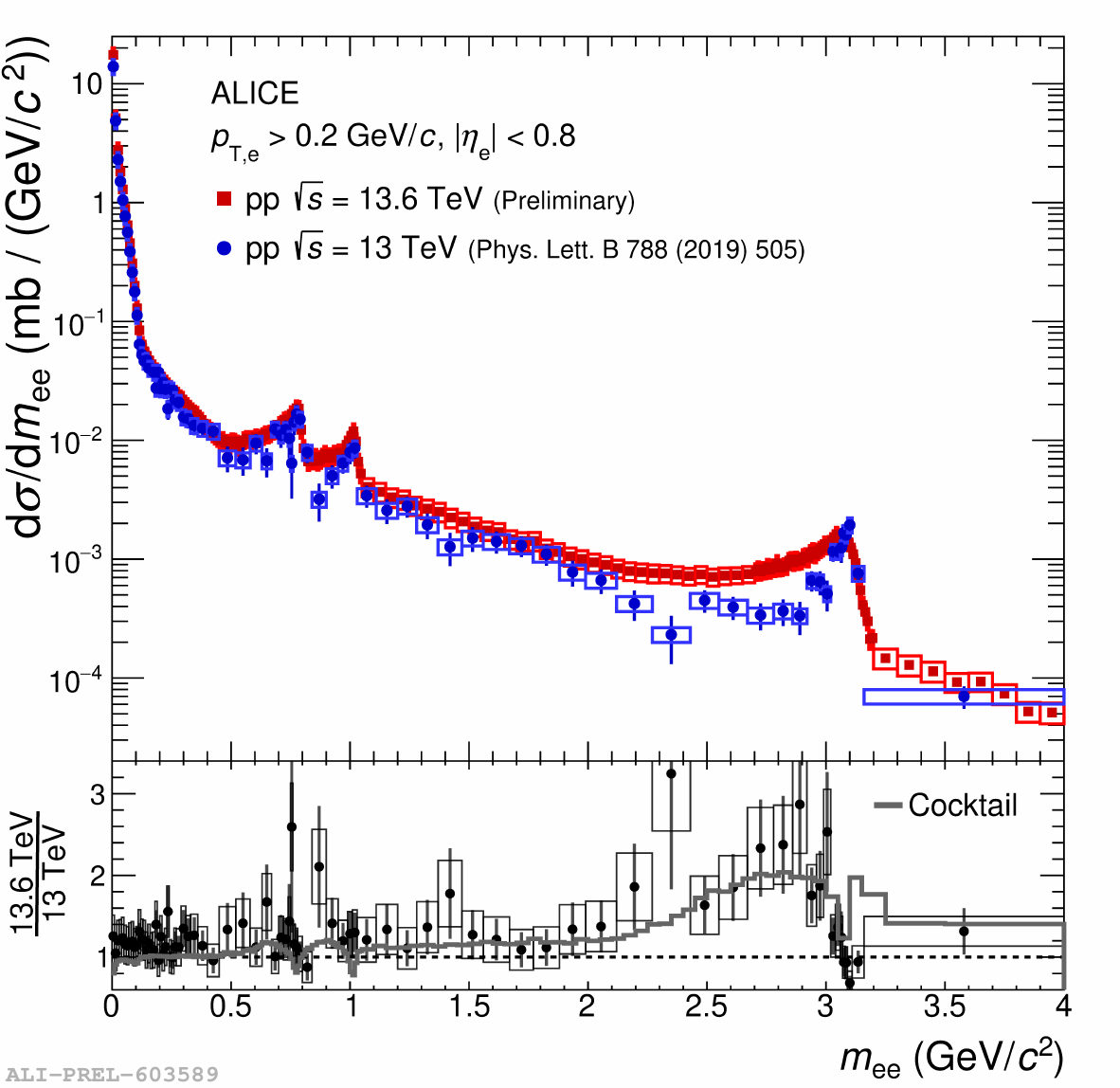}
\includegraphics[width=5cm,clip]{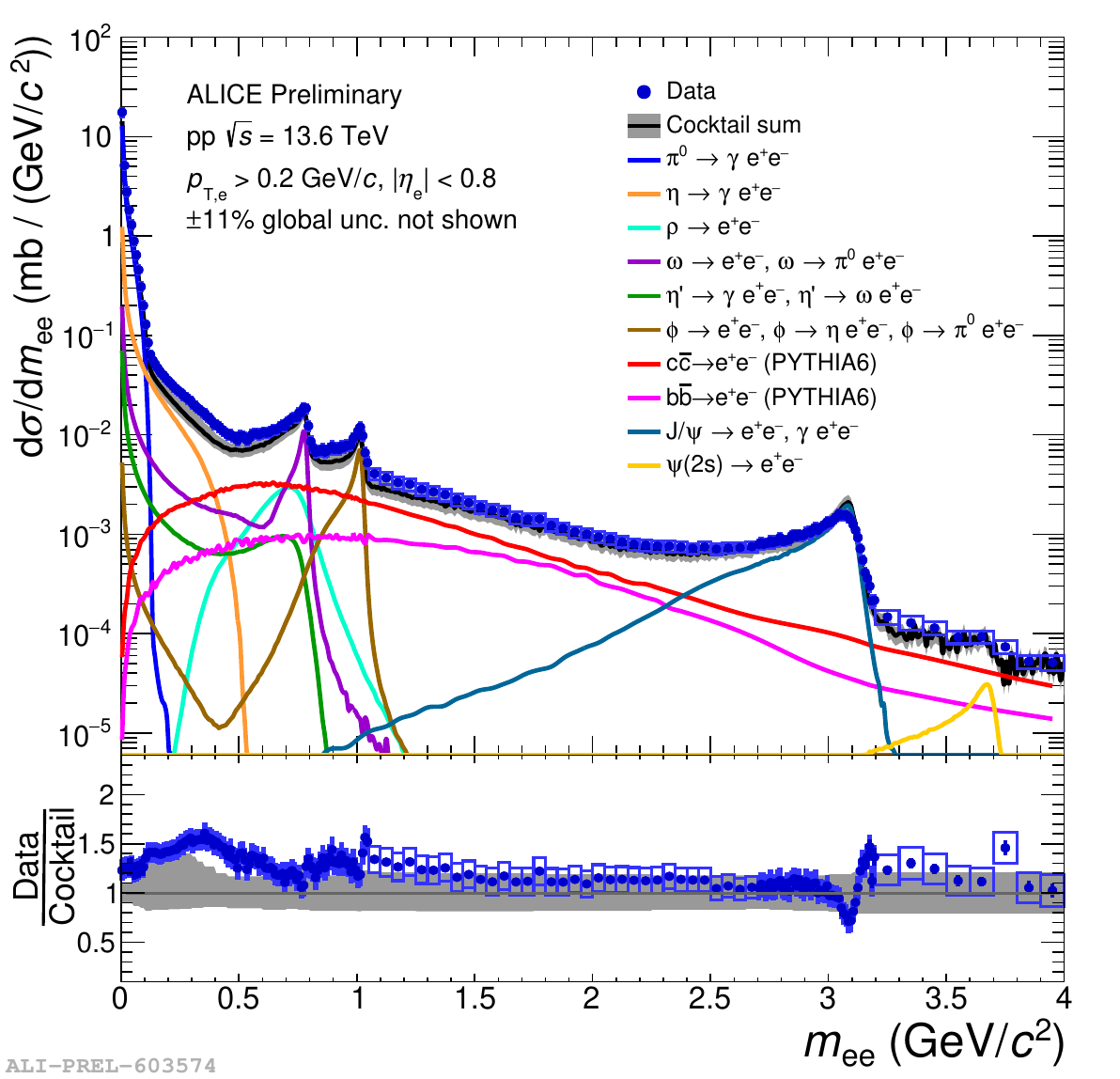}
\caption{Left: Comparison of dielectron production in pp collisions at \sqrtspp~ (red) and 13 TeV (blue). The bottom panel shows the ratio between both energies. Right: Dielectron cross section compared to the hadronic cocktail with the ratio data over cocktail. (Using 2023 data sample.)}
\label{Mee_run2run3_datacoktail}
\end{figure}

With the large data sample available in Run 3, dielectron measurements enter a new era of unprecedented precision.
%
%
% \autoref{rawSignal_pp} presents the raw invariant mass spectrum of the minimum-bias pp data collected in 2023. The signals shows the characteristic features of the $\pi^{0}$ and the resonances of $\omega$, $\phi$, J/$\psi$ and also the $\Upsilon$ at higher masses.
% \begin{figure}[h]
% % Use the relevant command for your figure-insertion program
% % to insert the figure file.
% \centering
% \includegraphics[width=5cm,clip]{macro-latex-web-conf/plots/raw_signal2023.pdf}
% \caption{Please write your figure caption here}
% \label{rawSignal_pp}
% \end{figure}
%
%
 The first measurement of the dielectron production cross section in pp collisions at \sqrtspp~ is shown in \autoref{Mee_run2run3_datacoktail} (left) together with the one at \mbox{$\sqrt{s}$ = 13 TeV} \cite{Run2pp13TeV}.  
 Within statistical and systematic uncertainties, the two results are consistent with each other. 
 %The increased yield in the Run 3 data in the mass region of the J/$\psi$ can be explained by an increased bremsstrahlung contribution due to additional material in front of the TPC. 
 % This effect is also described by Monte Carlo (MC) simulations for the hadronic cocktail, shown for reference as the grey line in the bottom panel.
 %
 %The increased yield in the Run 3 data to the left of the J/$\psi$ peak can be explained by an increased bremsstrahlung contribution, due to additional material between the ITS and TPC, together with some resolution effects. 
 %This effect is described by Monte Carlo (MC) simulations for the hadronic cocktail, shown as grey line in the bottom panel, although the detector resolution is still not perfectly reproduced in Run 3.
 %
 %The difference in the J/$\psi$ region can be explained by a different detector response. The left flank shows a larger bremsstrahlung contribution due to changes in the material budget, while the right flank indicates a peak broadening due to a slightly worse \pt~resolution of the current reconstruction.
 %
 The difference in the J/$\psi$ region can be explained by a different detector response. The left flank shows a larger bremsstrahlung contribution due to a different material budget, combined with a peak broadening due to a slightly worse \pt~resolution, visible on the right flank.
 Those effects are also described by Monte Carlo (MC) simulations for the hadronic cocktail, shown as grey line in the bottom panel.
The hadronic cocktail is based on measurements of light-flavour mesons, as well as \mbox{PYTHIA 6} calculations for the heavy-flavour components, fitted to the measured dielectron spectrum in pp collisions at $\sqrt{s}$ = 13 TeV \cite{Run2pp13TeV}. The energy dependence of the heavy-flavour production is estimated using FONLL calculations \cite{FONLLCalculation}. Moreover, the Run 3 detector resolution is applied. In \autoref{Mee_run2run3_datacoktail} (right), the hadronic cocktail and its components are compared to the measured dielectron cross-section at \sqrtspp. Within statistical and systematic uncertainties the data-to-cocktail ratio in the bottom panel shows a good agreement. In the resonance peaks, however, some residual resolution effects are observed. A tension of about 1.8 $\sigma$ is present in the mass region of the $\eta$ meson, for which a preliminary parametrization of the $\eta$/$\pi^0$ ratio was used in the cocktail.
\linebreak

\begin{figure}[h]
\centering
\includegraphics[width=5cm,clip]{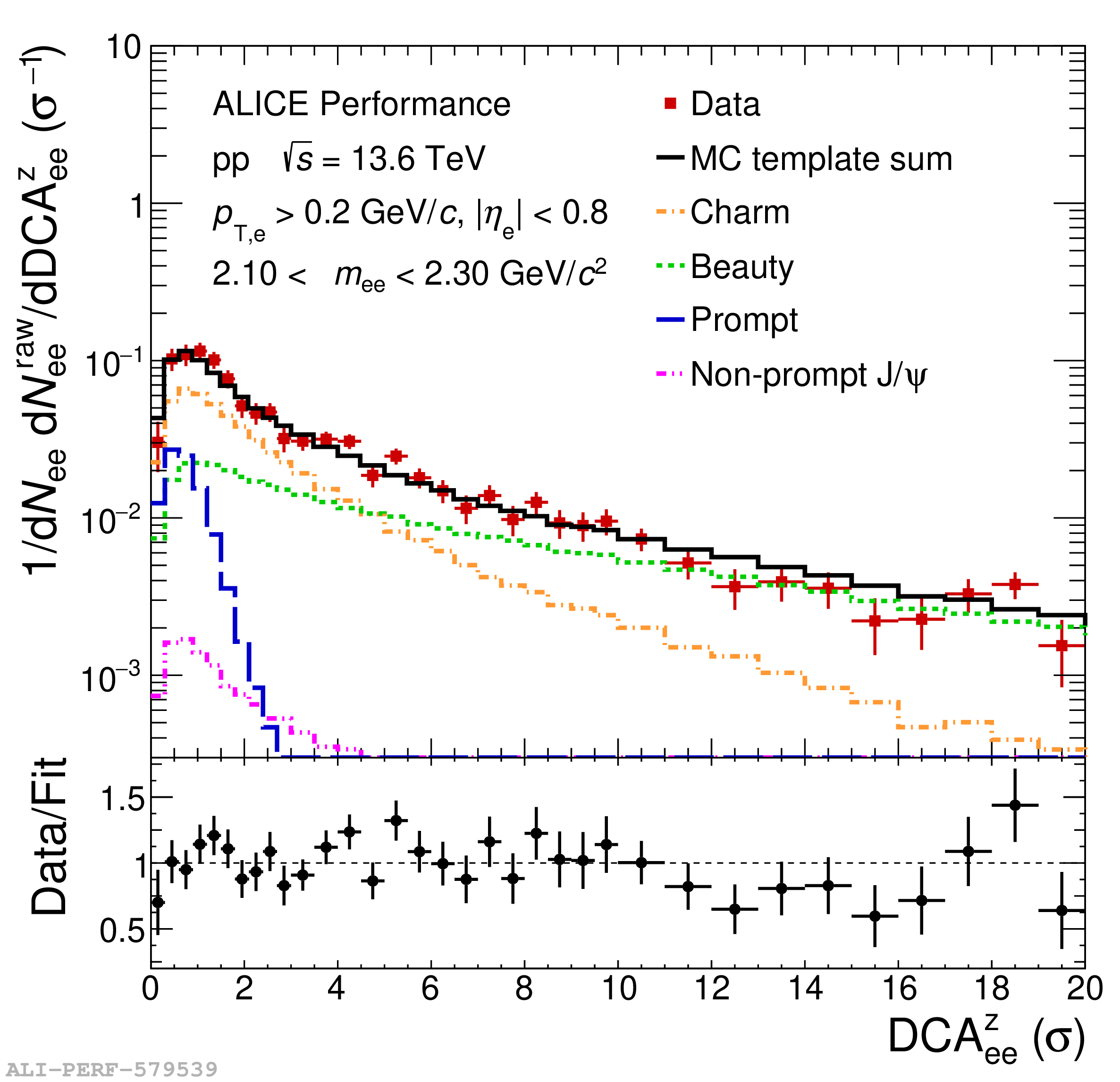}
\includegraphics[width=5cm,clip]{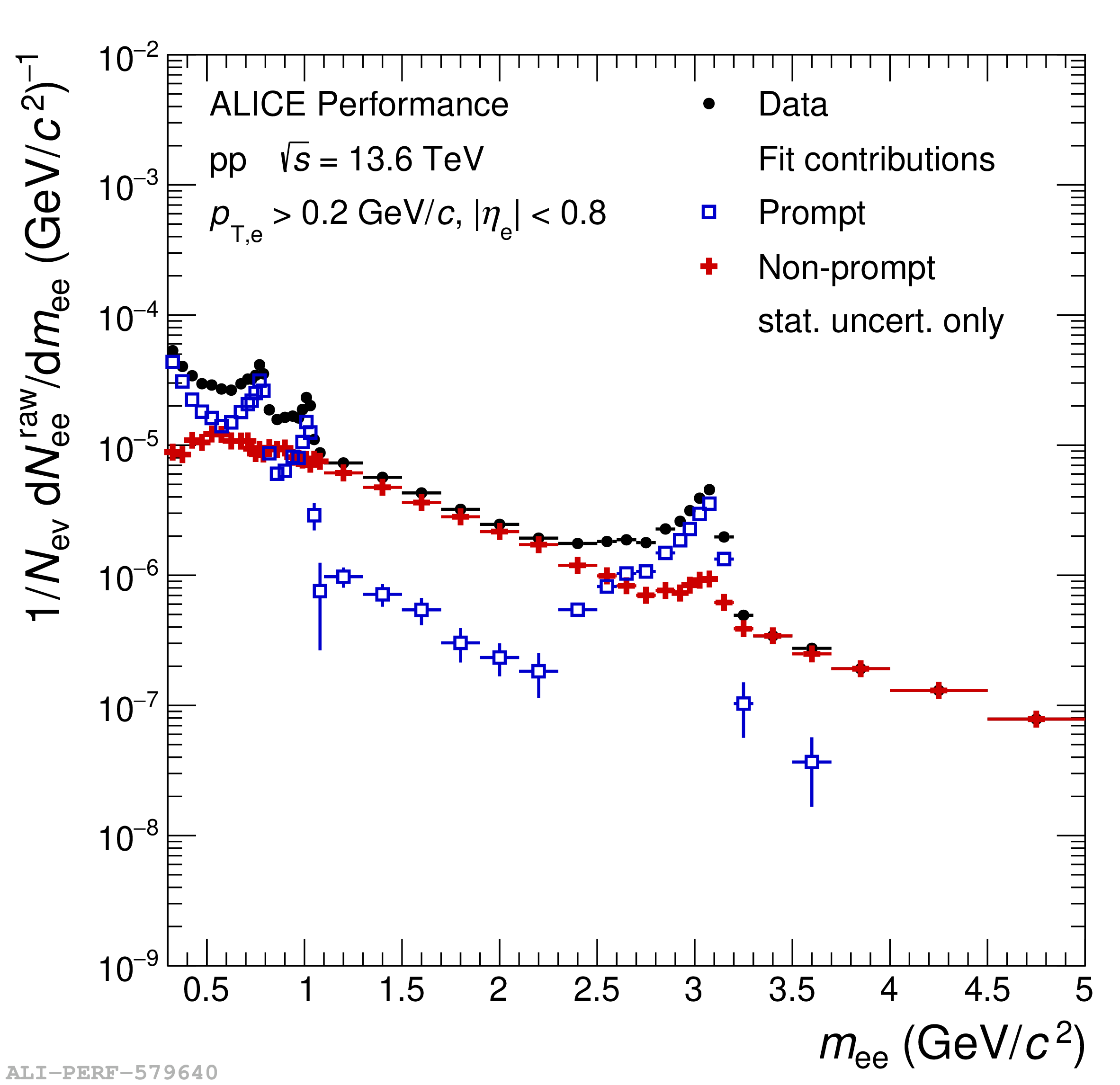}
\caption{Left: Raw dielectron yield as a function of \DCAeez~(\DCAee~ in the beam direction) measured in pp collisions at
$\sqrt{s} = 13.6$ TeV in the IMR fitted with different templates. Right: Corresponding unfolded invariant-mass spectra for prompt and non-prompt sources based on \DCAeez~template fits. (Using 2022 data sample.)}
\label{DCAfit_unfoldedMee}
\end{figure}

% \noindent
To estimate a contribution from the Drell--Yan process (and possible thermal radiation) in the IMR, the contributions of decays from charm and beauty mesons need to be handled. The decay length of heavy-flavour hadrons can be used to distinguish the HF background from prompt sources. With a decay length of the order $c\tau_{\rm{D}} \approx 150~\mu m$ and $c\tau_{\rm{B}} \approx 470~\mu m$ the particles travel a certain distance from the primary vertex before decaying weakly.
% \cite{PDGbook}.
Consequently, the electrons and positrons do not point directly towards the primary vertex. To distinguish these non-prompt dielectron decays from prompt sources, a pair variable the distance-of-closest approach (DCA) is defined as the quadratic mean of the DCA of each track normalised to its resolution $\sigma$, i.e. \mbox{DCA$_{\rm{ee}} = \sqrt{\frac{1}{2}\left[(\rm{DCA}_{e,1}/\sigma_1)^2+ (\rm{DCA}_{e,2}/\sigma_2)^2\right]}$ \cite{Run2pp7TeV, Run2PbPb502TeV}.} MC templates are created for each dielectron source and used to fit the measured raw \DCAee~ distribution in several mass intervals. 
The fit result, within the mass range of 2.1 < \mee ~< 2.3 \gevcc, over a large DCA range in z-direction is shown in \autoref{DCAfit_unfoldedMee} (left). The data-to-fit ratio is consistent with unity.
Due to their different \DCAee~ shapes, the prompt (light-flavour, Drell--Yan, prompt  J/$\psi$, possible thermal radiation) and non-prompt (heavy-flavour, non-prompt J/$\psi$) sources can be unfolded. This is presented in \autoref{DCAfit_unfoldedMee} (right). The raw non-prompt mass spectrum (red) shows the smooth continuum behaviour of the heavy-flavour decays, as well as a peak structure of non-prompt J/$\psi$ decays. In the prompt contribution (blue) the resonances are clearly visible. Looking at the IMR, it appears that the fit favours a small contribution from prompt sources. Such a signal is of great interest and will be the subject of further investigations.
\linebreak

\begin{figure}[h]
\centering
\includegraphics[width=5cm,clip]{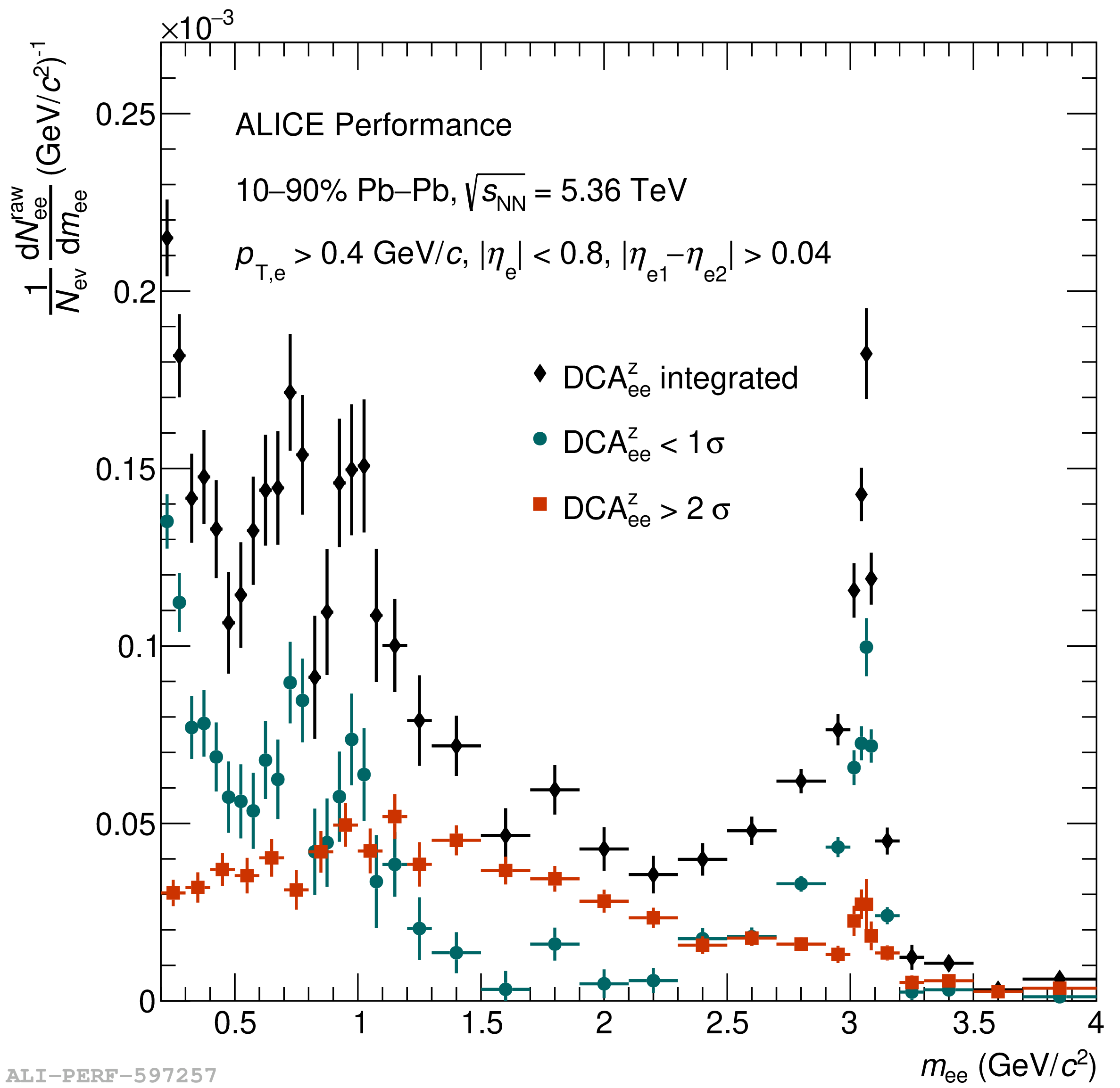}
\includegraphics[width=5cm,clip]{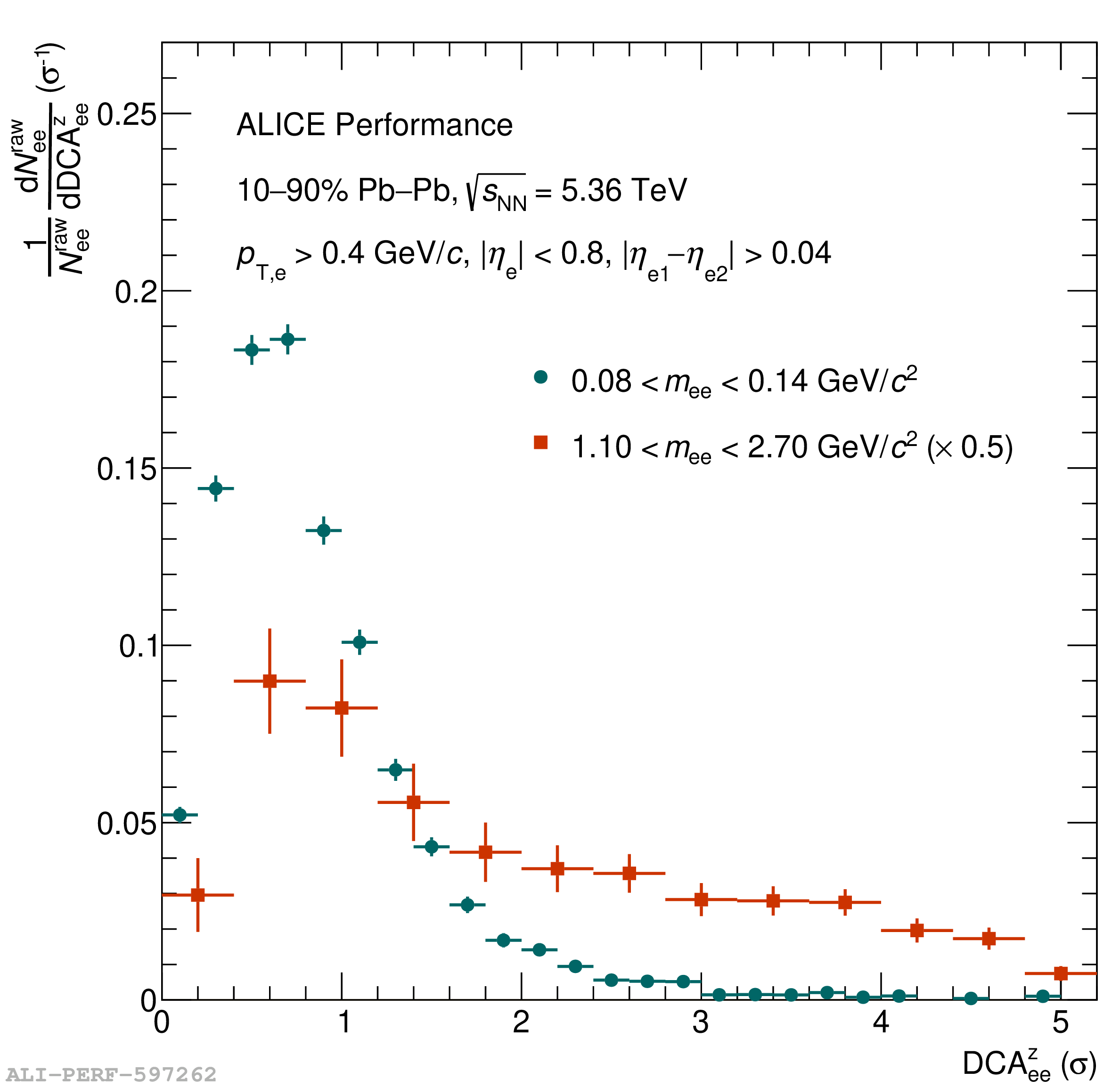}
\caption{Left: Raw dielectron signal as function of invariant mass, with selections of \DCAeez < 1$\sigma$ (dark cyan),  \mbox{\DCAeez > 2$\sigma$} (red) and the integrated \DCAeez~ (black) in Pb--Pb collisions at \mbox{\sqrtsPbPb}~for the centrality class 10-90$\%$ and a \pte~> 0.4 \gevc~ in Run 3. Right: Raw dielectron signal as function of \DCAeez~ selected in 0.08 < \mee < 0.14 \gevcc~(dark cyan) and 1.1 < \mee < 2.7 \gevcc~(red). 
(Using 2023 and 2024 data sample.)
}
\label{Mee_Signal_PbPbDCA3}
\end{figure}

%
% Pb--Pb
%
% \noindent
Compared to dielectron measurements in pp collisions, the measurements in Pb--Pb collisions are much more challenging. The multiplicity in each event is larger, as is the combinatorial background, leading to a much smaller signal-to-background ratio. 
%Thus, such measurements benefit greatly from the large statistics available in Run 3.
The raw spectrum as a function of \mee~ in Pb--Pb collisions at \sqrtsPbPb~ (black) is presented in \autoref{Mee_Signal_PbPbDCA3} (left) for the centrality class 10-90$\%$ and a \pte~> 0.4 \gevc. The spectrum shows the characteristic signature of the %$\pi^0$-Dalitz decay at low mass, as well as the
resonance peaks of the $\omega$, $\phi$ and J/$\psi$.
In Pb--Pb collisions, QGP thermal radiation is expected in the IMR. The \DCAee~variable is then crucial to separate this contribution from the non-prompt heavy-flavour background. In \autoref{Mee_Signal_PbPbDCA3} (right), two selections in \mee~are presented. They showcase the different \DCA~shapes of the prompt and HF contributions. In \autoref{Mee_Signal_PbPbDCA3} (left) the dark cyan markers with \mbox{\DCAeez~< 1$\sigma$} show the part of the spectrum dominated by prompt sources, whereas the red markers with \mbox{\DCAeez~> 2$\sigma$} show the continuum dominated by heavy-flavour decays, with the non-prompt J/$\psi$ peak. The distribution show the separation power of the \DCAeez.

\section{Conclusion and outlook}
\label{conclusion}

% The upgrades of the ALICE apparatus for Run 3 provide great improvements for dielectron measurements. 
The first measurement of the dielectron cross section in pp collisions at \sqrtspp~ is presented with \mbox{ALICE} in Run 3. The large amount of data enables a significant increase in statistical precision of dielectron measurements compared to Run 2. Furthermore, the improved pointing resolution allows an unprecedented separation power between prompt and non-prompt dielectron sources over a wide mass range. The concept of unfolding the mass distribution into a prompt and a non-prompt spectrum is established through template fits to the measured \DCAee~inclusive dielectron distributions. It is also demonstrated in Pb--Pb collisions, that prompt and non-prompt sources can be separated by applying selection criteria in \DCAee. 
The next step is to unfold the \DCAee~spectrum, following the example of the pp analysis. The ability to unfold the dielectron spectra marks a milestone in extracting the QGP thermal radiation in Pb--Pb collisions until the end of \mbox{Run 3 $\&$ 4}.

\bibliography{References.bib} % Replace "your_bib_file" with the actual name of your .bib file
% \begin{thebibliography}{}
% %
% % and use \bibitem to create references.
% % Format for Journal Reference
% Journal Author, Article title. Journal \textbf{Volume}, page numbers (year). \url{https://doi.org/Article-DOI-number}
% Book Author, \textit{Book title} (Publisher, place, year) page numbers
% \end{thebibliography}

\end{document}